\documentclass[preprint,amsmath,amssymb,prb,groupedaddress]{revtex4-2}
\usepackage{graphicx}
\usepackage{dcolumn}
\usepackage{bm}
\usepackage[columnwise]{lineno}
\usepackage{CJKutf8}
\usepackage{hyperref}
\usepackage{color}

\begin{document}


%
%
%
\title{Positive and negative chemical pressure effects investigated in electron-doped FeSe films with an electric-double-layer structure}


\newcommand{\Tc}{$T_{\mathrm c}$ }
\newcommand{\Ts}{$T_{\mathrm {nem}}$ }
\newcommand{\Ig}{$I_{\mathrm g}$ }
\newcommand{\Vg}{$V_{\mathrm g}$ }
\newcommand{\FeSeS}{FeSe$_{1-y}$S$_y$ }
\newcommand{\FeSeTe}{FeSe$_{1-x}$Te$_x$ }
\newcommand{\Tczero}{$T_{\mathrm c}^{\mathrm {zero}}$ }
\newcommand{\Tconset}{$T_{\mathrm c}^{\mathrm {onset}}$ }
\author{N. Shikama}
\author{Y. Sakishita}
\author{F. Nabeshima}
\email[]{cnabeshima@g.ecc.u-tokyo.ac.jp}
\author{A. Maeda}
\affiliation{Department of Basic Science, the University of Tokyo, Meguro, Tokyo 153-8902, Japan}
\date{\today}

\begin{abstract}

We investigated chemical pressure effects in electron-doped (e-doped) FeSe by fabricating an electric-double-layer structure with single crystalline FeSe films on LaAlO$_3$ with Se substituted by isovalent Te and S.    Our method enables transport measurements of e-doped FeSe.   Electron doping by applying gate voltage of 5 V increases \Tc of the FeSe$_{1-x}$Te$_x$ and FeSe$_{1-y}$S$_y$ films with $0 \leq x \leq 0.4$ and $0 \leq y \leq 0.25$, while the e-doped $x=0.5$ film showed lower \Tc than that of the undoped one.      Both positive and chemical pressure suppress \Tc of the e-doped FeSe.    The obtained superconducting phase diagram in the e-doped samples is rather different from that in undoped samples.   This might suggest that the superconductivity mechanism is different between undoped and e-doped systems.   Alternatively, this is possibly explained by the absence of the nematic order in e-doped samples.

\end{abstract}

\maketitle
\newpage
\section{Introduction}
Iron chalcogenide superconductor, FeSe\cite{Wu08}, has attracted much attention  in the field of superconductivity.     FeSe shows a wide variety of intriguing physical properties\cite{JPCM.30.023001,ARCMP.9.125,FeSeReview-Kreisel-2020,JPSJ.89.102002}such as the nematic state apart from magnetic order, topological superconductivity, and other exotic superconducting states.   FeSe also shows drastic enhancement of the superconducting transition temperature, $T_{\mathrm c}$, in various ways.   The \Tc value of FeSe under ambient pressure is approximately 9 K, but it increases up to 40-50 K by applying hydrostatic pressure\cite{NatMat.8.630,Marga09} and by doping electron through intercalation\cite{PhysRevB.82.180520,SciRep.2.426,PhysRevB.92.064515,Shi_2018} and the electric-field effect\cite{PRL.116.077002,PNAS.113.3986,NatPhys.12.42}.  In addition, a large gap was observed in monolayer FeSe films on SrTiO$_3$ (STO)\cite{CPL.29.037402}.    The following angle-resolved photoemission spectroscopy (ARPES) measurements revealed that the gap opens below 65 K\cite{NatMater.12.605,NatMater.12.634}, which was argued to be the manifestation of superconductivity.

While undoped FeSe has both electron and hole Fermi surfaces like other iron-based superconductors, electron-doped (e-doped) FeSe with enhanced \Tc as high as 40-50 K has electron Fermi surfaces alone\cite{PhysRevB.92.060504}.  Monolayer FeSe on STO has the similar electronic state to e-doped FeSe by charge transfer from the substrate\cite{NatCommun.3.931}.  An interesting question is whether the mechanism of superconductivity in the heavily e-doped FeSe is the same as that of undoped FeSe.  Another important issue is if there is any interface effect between the film and the substrate which enhances the superconductivity in the monolayer FeSe on STO.  Indeed, resistive transition at as high as 65 K has not been confirmed in monolayer FeSe, below which the energy gap was observed by ARPES.  Although Ge $et\ al.$ reported zero resistivity at a very high temperature of approximately 100 K\cite{Ge_NatMater.14.285}, its reliability and reproducibility have been controversial\cite{BozovicNPhys2014}.   Rather, recent $in$-$situ$ resistivity measurements reported that the highest temperature where the zero resistivity was observed, $T_{\mathrm c}^{\mathrm {zero}}$, was approximately 30 K\cite{PhysRevLett.124.227002,arXiv:2010.11984}, much lower than the gap-opening temperature and also lower than \Tczero of the e-doped FeSe.    It is also suggested that the energy gap at high temperatures observed by the ARPES measurements in ultra-thin samples is a pseudo-gap due to the strong superconducting fluctuations\cite{arXiv:2010.11984}.   To tackle these problems, systematic investigation of the physical properties in e-doped FeSe is of particular importance. 

Isovalent Te- and S-substitution for Se is extensively studied in undoped FeSe.  Te- and S-substitution expands and shrinks the lattice of FeSe, respectively, without doping carriers, and thus is referred to as the positive and negative chemical pressure, respectively.    However, research on electron doping so far focused only on FeSe and there are few reports on element-substitution effects for single crystalline samples.   
Although some papers reported intercalation experiments on powder samples of FeSe$_{1-x}$Te$_x$ and FeSe$_{1-y}$S$_y$\cite{HLei_PhysRevB.90.214508,PhysRevB.96.064512,PhysRevB.96.134503,JGuo_NatCom.5.4756,XFLu_PhysRevB.90.214520}, it is difficult to measure transport properties with powder samples\cite{JPSJ.83.113704,JPSJ.85.013702}.    

In this paper, we report fabrication of the electric-double layer transistor (EDLT) structure with single crystalline thin films of FeSe$_{1-x}$Te$_x$ and FeSe$_{1-y}$S$_y$, which are available for a wide range of composition\cite{yi15pnas,yiSciRep17,Nabe18.FeSeS}.   Our method using epitaxial films enables transport measurements of e-doped samples.   We investigated the chemical pressure effect on the superconductivity of e-doped FeSe$_{1-x}$Te$_x$ and FeSe$_{1-y}$S$_y$ thin films for $x$ $\leq$ 0.25, and $y$ $\leq$ 0.5.   All the films except $y$ = 0.5 showed the increase of \Tc after the electron doping.    \Tc of the electron-doped FeSe films decreases with both Te and S substitution.     The obtained superconducting phase diagram of electron-doped FeSe$_{1-y}$S$_y$ and FeSe$_{1-x}$Te$_x$ films is rather different from that of the undoped films and that of monolayer FeSe$_{1-x}$Te$_x$ films.      This might suggest that the superconductivity mechanism is different between undoped and e-doped systems.   Alternatively, this is possibly explained by the absence of the nematic order in e-doped samples.

\section{Method}
All the films were grown on LaAlO$_3$ (LAO) substrates using a pulsed laser deposition method, details of which are described elsewhere\cite{Imai09,Imai10}.    The EDLT structure was fabricated on the grown films with DEME-TFSI as the gate dielectric \cite{PRL.116.077002,PNAS.113.3986,NatPhys.12.42,Kouno18,ShikamaAPEX2020}.   Once the sample is taken out from the growth chamber to the atmosphere, the surface of the films is degraded and becomes inactive to the electrostatic doping.   To obtain enhanced \Tc by electron doping, we removed the surface dead layer by the {\it in-situ} electrochemical etching of the film with the ionic liquid\cite{Kouno18}.     The gate voltage, $V_{\mathrm g}$, of 5 V, was applied at 220 K, and then the temperature was raised and kept at around 240 K to induce etching.    Details of the etching process are described in our previous papers\cite{Kouno18,ShikamaAPEX2020}.   We found that an e-doped layer is formed at the surface of the film during the etching process by an electrochemical reaction, possibly by the intercalation of DEME$^+$ cations in FeSe or by FeSe which adsorbs DEME$^+$ cations on the surface, and that the reacted layer exhibits the high \Tc in the FeSe-EDLT systems\cite{Kouno18,ShikamaAPEX2020}.   A single etching process was finished when the film thickness was decreased by approximately a few nanometers.   We estimated the thickness of the etched films from the time integral of the gate current.   We repeated the cycle of an etching process and an accompanying measurement of the temperature dependence of the resistivity.      The electrochemical etching and the resistivity measurements were performed in a helium atmosphere using a Quantum Design Physical Property Measurement System.    

\section{Results and Discussion}

\begin{figure*}[hbtp]
 \centering
 \includegraphics[width=\linewidth,clip]
      {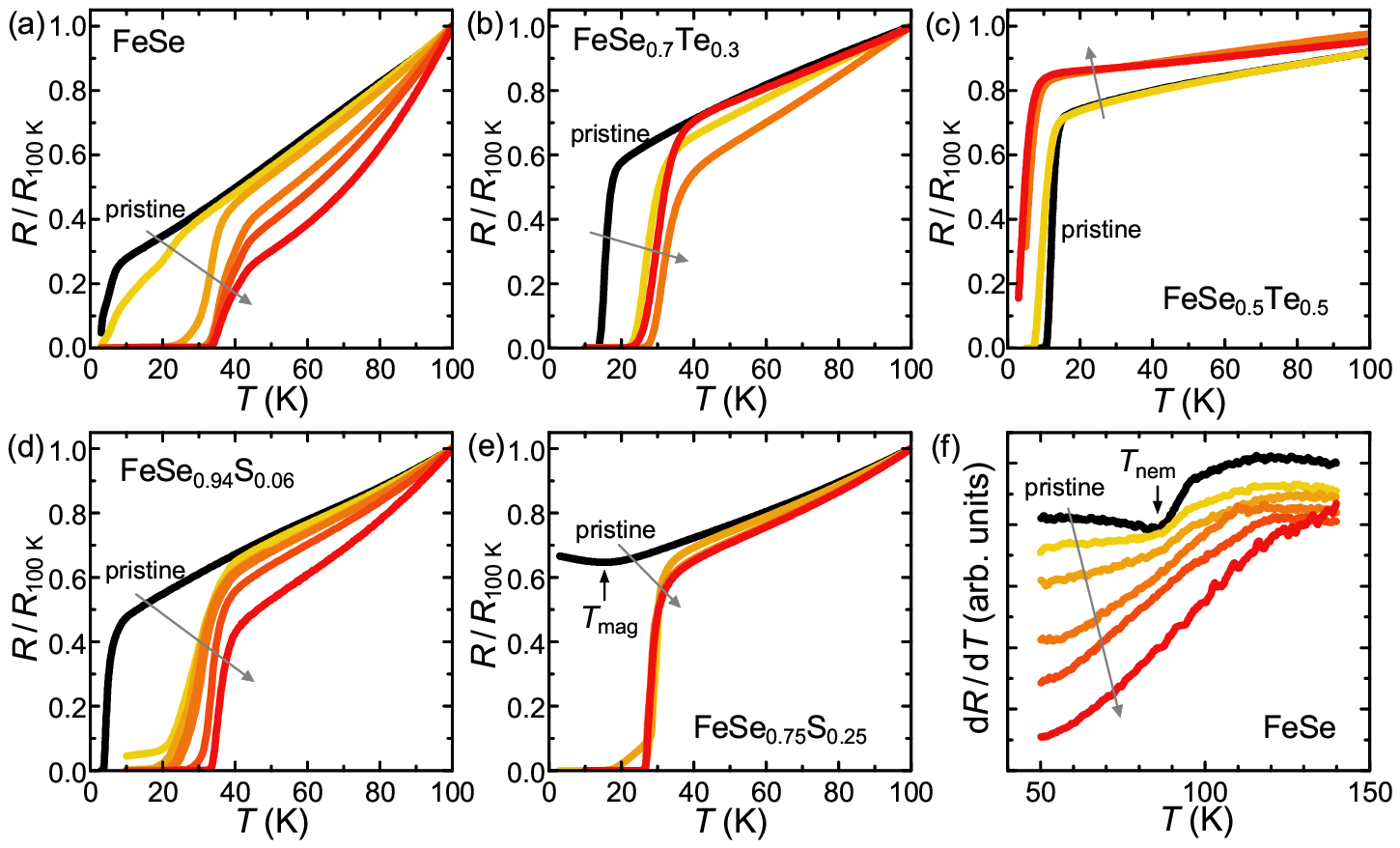}
 \caption{(a)-(e) Shift of the temperature dependence of the resistance, $R$ of the etched FeSe$_{1-x}$Te$_{x}$ and FeSe$_{1-y}$S$_{y}$ thin films as the etching process was repeated for $x$ = $y$ = 0, $y$ = 0.3, $y$ = 0.5, $y$ = 0.06, and $y$ = 0.25, respectively. Arrows indicate the shift over the etching processes. (f) The temperature dependence of the temperature derivative of resistance of the etched FeSe. } 
\label{EtRT}
\end{figure*}

Figure \ref{EtRT}(a)-(e) shows the shifts in the temperature dependence of the resistance ($R$-$T$) of the electrochemically etched FeSe$_{1-x}$Te$_x$ and FeSe$_{1-y}$S$_y$ films when the etching process were repeated.    The $R$-$T$ curves and \Tc values change as the etching process is repeated.     The change of the \Tc values saturates after repeating the etching several times, as we previously reported in FeSe and FeSe$_{0.8}$Te$_{0.2}$ \cite{Kouno18,ShikamaAPEX2020}.   The $R$-$T$ curve of the undoped sample for $y=0.25$ showed an up-turn behavior at low temperatures.    This is due to a magnetic transition, which is characteristic in strained film samples of FeSe$_{1-y}$S$_y$\cite{Nabe.submitted}.   Because a single etching process increased \Tc up to higher temperatures than the magnetic transition temperature in the present study, the behavior of the magnetic phase over electron doping is not clear.    A systematic study of changing doping amount in FeSe$_{1-y}$S$_y$ would provide new insights into the relation between the superconductivity in e-doped FeSe and the magnetism, which is now underway.   

The change in $R$-$T$ of FeSe$_{0.5}$Te$_{0.5}$ thin films over electrochemical etching is different from that of other FeSe$_{1-x}$Te$_x$ and FeSe$_{1-y}$S$_y$ films.   As the etching cycle was repeated, the temperature dependence of resistance became weaker, and \Tc was suppressed gradually.   We naively expect that the e-doped layer near the surface of the film is responsible for the suppressed $T_{\mathrm c}$.   This seems, however, strange because of the following reason.   If the surface e-doped layer has lower $T_{\mathrm c}$, the supercurrent should flow in the underlying undoped layer with higher $T_{\mathrm c}$ and \Tc observed by the resistance measurements should be unchanged.   However, the suppression of the superconductivity by electron doping was also reported in bulk FeSe$_{0.5}$Te$_{0.5}$ with a field-effect transistor structure with a solid ion conductor\cite{PhysRevB.95.174513}.    Thus, it is likely that the electron doping really suppresses the superconductivity in the $x$ = 0.5 film.   In other words, the e-doped reacted layer might be formed in almost whole sample, not only near the surface, for the $x$ = 0.5 film.


Figure \ref{EtRT}(f) shows the temperature dependence of the temperature derivative of the resistance, $\mathrm d R/ \mathrm d T$, of the etched FeSe film.    It is well-known that an anomaly was observed in the temperature dependence of the resistivity at the nematic transition of FeSe.    The undoped FeSe film shows a dip anomaly at around 85 K in $\mathrm d R/ \mathrm d T$.    The dip anomaly became obscured as the etching process was repeated, which suggests the suppression of the nematic transition by electron doping.   Indeed, for FeSe flakes with an EDLT structure, the nematic transition is suppressed almost completely by applying \Vg = 4 V \cite{CPL.33.057401}.    Suppression of the nematic transition was also observed for a K-dosed thick FeSe film by an ARPES measurement\cite{NatCommun.7.10840}.

\begin{figure*}[hbtp]
 \centering
 \includegraphics[bb=0 5 1088 847,width=\linewidth,clip]
       {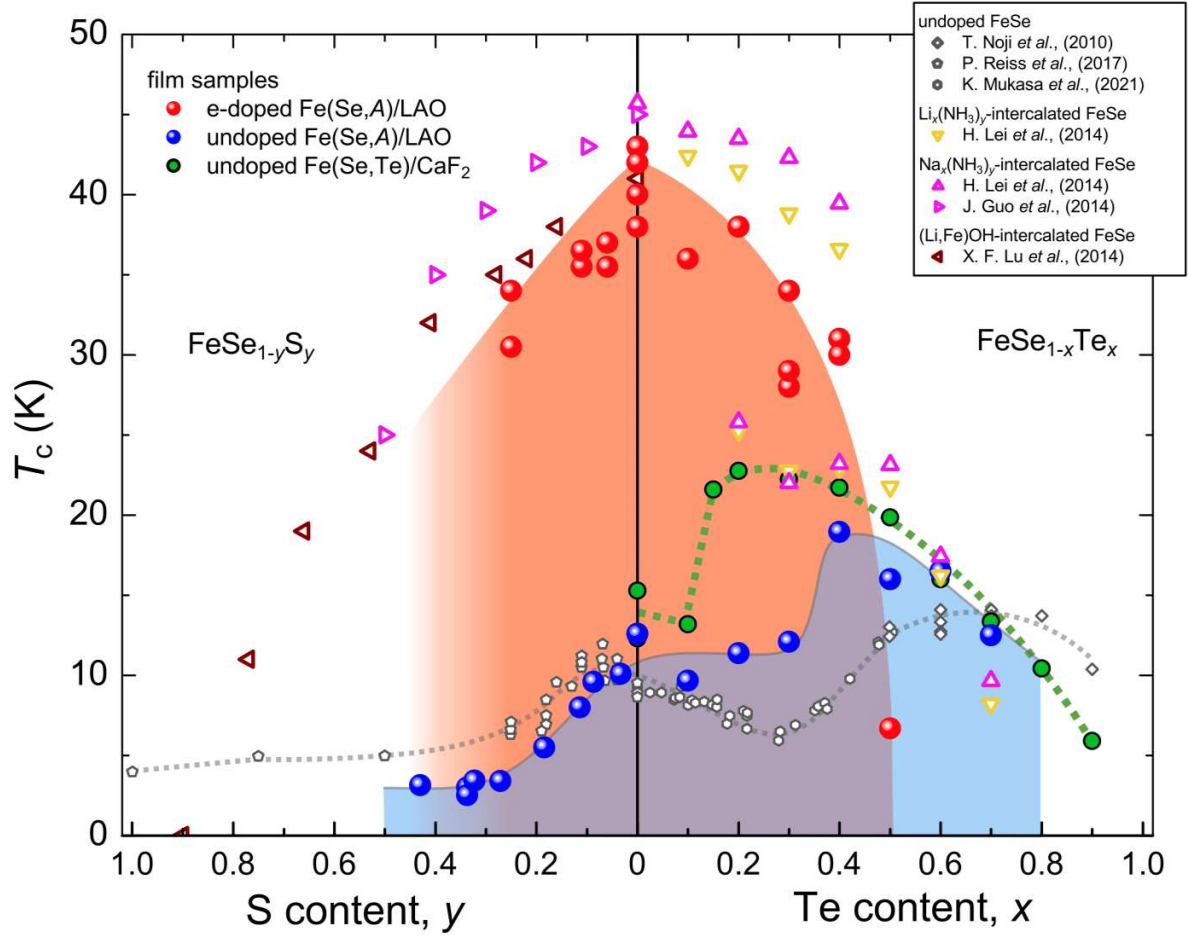}
  \caption{The Te- and S-content dependence of \Tc in undoped\cite{yiSciRep17,Nabe18.FeSeS} and e-doped FeSe$_{1-x}A_x$ ($A$=Te, S) films on LAO. Those of undoped FeSe$_{1-x}$Te$_x$ films on CaF$_2$\cite{yi15pnas,yi16pc} and those of bulk undoped\cite{JPSJ.79.084711,PRB.96.121103,Mukasa21_NCom} and intercalated samples\cite{HLei_PhysRevB.90.214508,JGuo_NatCom.5.4756,XFLu_PhysRevB.90.214520} are also plotted. The dotted lines are guides for the eye.}
 \label{PD2D}
 \end{figure*}

Figure \ref{PD2D} shows the composition dependence of \Tc of the electrochemically etched FeSe$_{1-x}$Te$_x$ and FeSe$_{1-y}$S$_y$.   \Tc of the e-doped films systematically decrease with increasing Te and S content, including the $x=0.5$ film which shows lower \Tc than the undoped sample.    For comparison, the data for powder samples of Li$_x$(NH$_3$)$_y$-\cite{HLei_PhysRevB.90.214508}, Na$_x$(NH$_3$)$_y$-\cite{HLei_PhysRevB.90.214508,JGuo_NatCom.5.4756}, (Li,Fe)OH-intercalated\cite{XFLu_PhysRevB.90.214520} Fe(Se,$A$) ($A$ = Te, S), and undoped FeSe$_{1-x}$Te$_x$ films on CaF$2$\cite{yi15pnas,yi16pc} are also shown in Fig. \ref{PD2D}.
Our results resembles those of bulk powder samples of intercalated FeSe$_{1-x}$Te$_x$ and FeSe$_{1-y}$S$_y$. 
Small variations in the \Tc values among these systems will be due to the difference of the intercalated ions.

As for the undoped samples, the composition dependence of \Tc is different among bulk and films on LAO and CaF$_2$, particularly for FeSe$_{1-x}$Te$_x$.     We attribute the difference to the difference in lattice strain in the samples.   It should be noted that films on CaF$_2$ and LAO are under strong and weak compressive strain, respectively, at least for Te-substituted samples.   \Tc of FeSe$_{0.5}$Te$_{0.5}$ increases with increasing compressive strain, and \Tc of a film on CaF$_2$ is approximately 1.5 times higher than that of bulk.    In addition, for samples under stronger compressive strain, the highest \Tc is achieved at smaller $x$ and the change in \Tc in the small $x$ side of the superconducting dome become steeper.      Thus, the composition dependence of \Tc seems to change systematically with the strength of the strain.

\begin{figure*}[hbtp]
 \centering
 \includegraphics[width=\linewidth,clip]
      {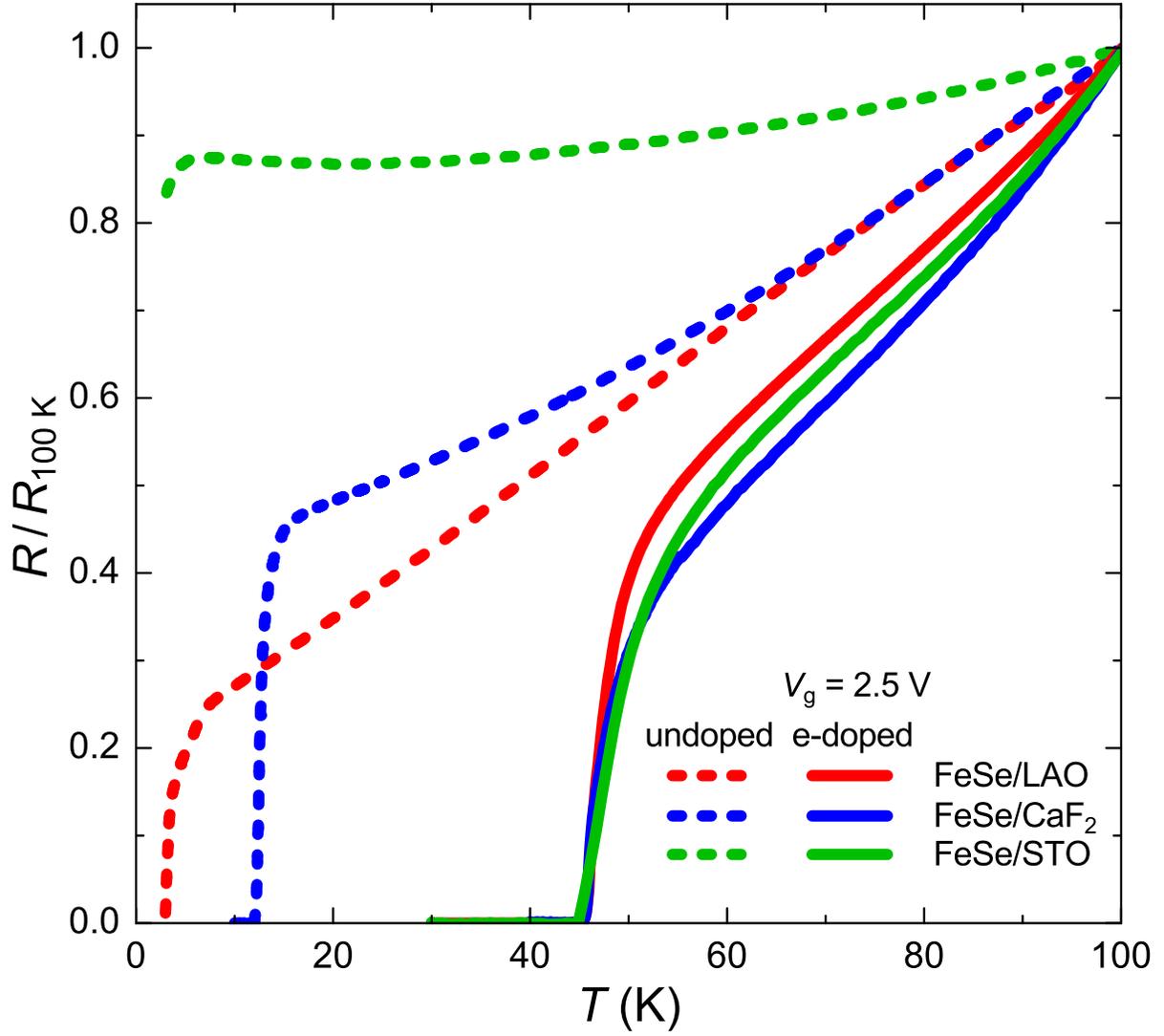}
 \caption{Temperature dependence of the normalized resistance of FeSe films on LAO, CaF$_2$, and STO substrates with an EDLT structure before and after the electrochemical etching/doping with \Vg = 2.5 V.} 
\label{EtRT_dSub}
\end{figure*}

Compared with the undoped samples, differences in the composition dependence of \Tc are much smaller between bulk and the films for the e-doped samples.   In addition, we previously observed almost the same \Tc in the e-doped FeSe$_{0.8}$Te$_{0.2}$ films on different substrates\cite{Kouno18}.    Figure \ref{EtRT_dSub} shows the temperature dependence of resistance of FeSe films on LAO, CaF$_2$ and STO substrates with an EDLT structure before and after the electrochemical etching/doping with \Vg = 2.5 V.     Despite the different $T_{\mathrm c}$'s of the undoped films, which are likely due to the different strength of the strain, the e-doped films shows exactly the same \Tc values (\Tczero = 46 K).    These results show that \Tc does not depend on the strain in e-doped films, which is in contrast to the undoped samples.     There are two possible explanations for the similar \Tc in e-doped samples.    One explanation is that the lattice strain does not affect the superconductivity in e-doped FeSe.   The other possibility is that the strain in the e-doped films is relaxed at least in the reaction layer at the surface with high $T_{\mathrm c}$.    Considering the fact that \Tc of intercalated FeSe largely changes by applying physical pressure\cite{Shahi_PRB2018,JPSun_NCom.9.380} and also by applying chemical pressure, it is unlikely that the strain would not change \Tc of the e-doped FeSe.   Thus, the strain would be relaxed in the reacted layer with high $T_{\mathrm c}$.


While the undoped films shows a significant change in \Tc at $x$ = 0.3-0.4, the composition dependence of \Tc in the e-doped films shows a simple dome-like behavior with the peak at $x=y=0$.    This difference may suggest different superconducting mechanisms between undoped and e-doped FeSe.    Alternatively, it can be also explained by the nematic order in undoped FeSe.    The composition where the significant change in \Tc is observed in undoped FeSe$_{1-x}$Te$_x$ coincides with the disappearance of the nematic order\cite{yiSciRep17,Nakayama21_PRR3.L012007}.   A Magneto-transport experiment revealed a correlation between \Tc and carrier density\cite{Nabe20_PRB}, suggesting that the band reconstruction in the nematic phase predominantly affects the superconductivity rather than the existence of the nematic order itself or the nematic fluctuation possibly developing near the quantum critical point.    Indeed, our recent DFT calculation showed a drastic change in DOS at the Fermi level before and after the nematic transition\cite{Kurokawa.submitted}.     In contrast to the undoped FeSe, the e-doped FeSe does not exhibit the nematic order, as was suggested by the resistance measurement.    Thus, the different composition dependence of \Tc can be explained by the presence/absence of the nematic order in these two systems.

\begin{figure*}[hbtp]
 \centering
 \includegraphics[bb=0 0 1140 830,width=\linewidth,clip] 
      {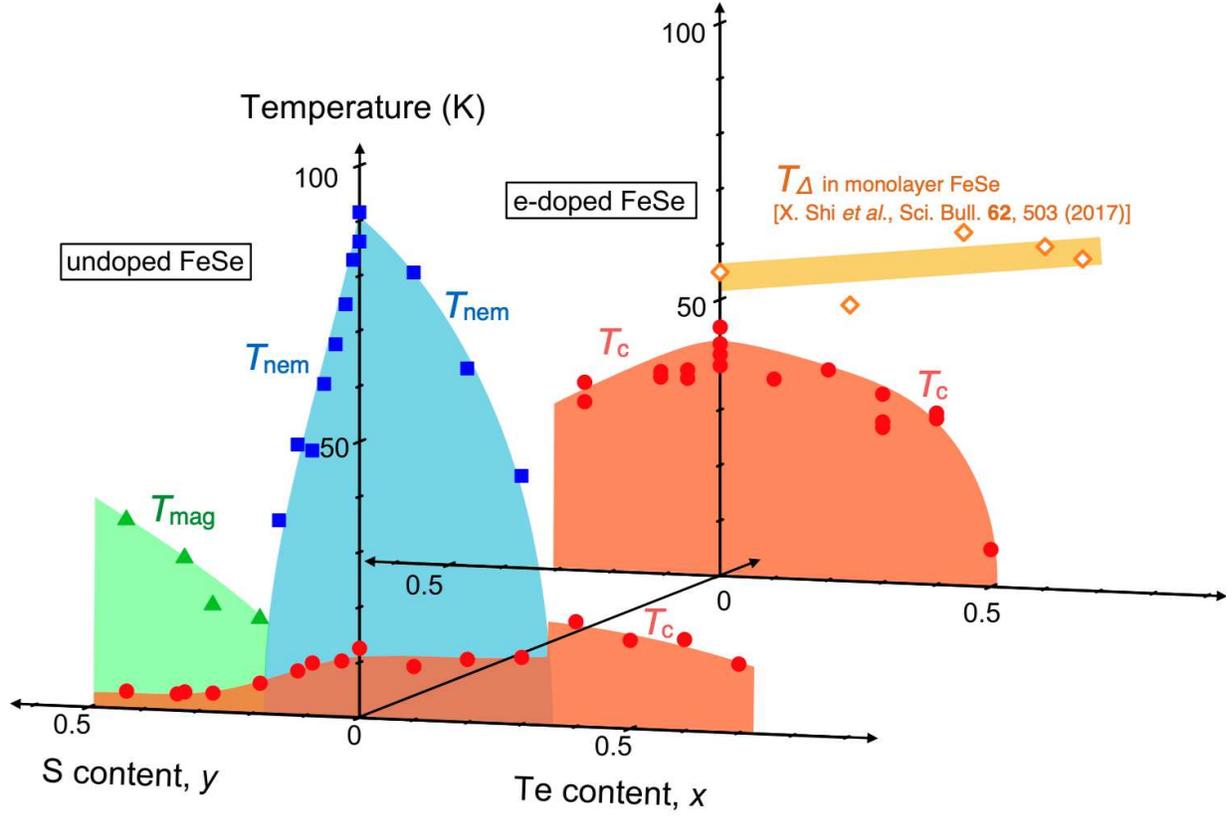}
 \caption{Obtained phase diagram of the undoped\cite{yiSciRep17,Nabe18.FeSeS} and the e-doped FeSe$_{1-x}$Te$_{x}$ and FeSe$_{1-y}$S$_{y}$ films. $T_{\mathrm {mag}}$ represents the magnetic transition temperature in undoped FeSe$_{1-y}$S$_{y}$ films\cite{Nabe.submitted}. The gap-opening-temperature, $T_\Delta$, of the single-layer FeSe$_{1-x}$Te$_{x}$/SrTiO$_3$\cite{Shi17.SciBull.62.503} are also plotted.}
\label{PD3D}
\end{figure*}

Figure \ref{PD3D} shows the phase diagram of e-doped and undoped FeSe thin films.   The gap-opening temperature, $T_\Delta$, for monolayer FeSe$_{1-x}$Te$_x$ films on STO are also plotted.    $T_\Delta$ of the monolayer FeSe$_{1-x}$Te$_x$ films is almost independent of $x$, while our e-doped films of FeSe$_{1-x}$Te$_x$ shows the monotonic decrease of $T_{\mathrm c}$.   This might suggest that the superconductivity mechanism of monolayer FeSe is different from that of e-doped FeSe.   In fact, theoretical studies suggested that the interfacial effect enhances the superconductivity in monolayer FeSe\cite{DHLee_CPB2015}.   Another possible scenario is that the energy gap observed by the ARPES studies in monolayer FeSe is not the superconducting gap and the true values of \Tc at which resistive transition occurs in monolayer FeSe is similar to that of e-doped FeSe.    A recent combined study of ARPES and resistivity measurements on monolayer FeSe revealed that the temperature where the resistivity start to decrease coincides with the gap-opening temperature, suggesting the opening of a pseudo-gap at the temperature due to strong superconducting fluctuations\cite{arXiv:2010.11984}.   A pseudo-gap behavior was also reported in organic-ion-intercalated FeSe by a nuclear magnetic resonance measurement\cite{Kang20_PRL125.097003}.    To settle this problem, investigation of the superconducting fluctuations in e-doped FeSe$_{1-x}$Te$_x$ is of great importance.   Transport measurements of the e-doped FeSe$_{1-x}$Te$_x$, combined with a study of the Meissner screening by a mutual inductance measurements will provide new insights into the superconductivity in FeSe-related materials, which is under way.

\section{Conclusion}
We investigated the chemical pressure effect of single crystalline thin films of FeSe$_{1-y}$S$_y$ and FeSe$_{1-x}$Te$_x$ by fabricating an EDLT structure, and obtained the phase diagram of e-doped FeSe$_{1-x}$Te$_x$ and FeSe$_{1-y}$S$_y$ films. Our method enables transport measurements in e-doped FeSe$_{1-x}$Te$_x$ and FeSe$_{1-y}$S$_y$.    \Tc of almost all the FeSe$_{1-x}$Te$_x$ and FeSe$_{1-y}$S$_y$ films was increased, while that of the FeSe$_{0.5}$Te$_{0.5}$ film was decreased. \Tc of the electron-doped films decreases monotonically by the substitution Se for S and Te.    This is rather different from those in undoped samples.     This might suggest that the superconductivity mechanism is different between undoped and e-doped systems.   Alternatively, this is possibly explained by the absence of the nematic order in e-doped samples.

\begin{acknowledgments}
This research was supported by JSPS KAKENHI Grant Numbers 18H04212 and 19K14651 and by the Precise Measurement Technology Promotion Foundation (PMTP-F).
\end{acknowledgments}



%

\newpage


\end{document}